\documentclass[%
 aip,
 amsmath,amssymb,
 reprint,%
fleqn]{revtex4-2}

\usepackage{graphicx}% Include figure files
\usepackage{dcolumn}% Align table columns on decimal point
\usepackage{bm}% bold math
\usepackage{color}

\usepackage[utf8]{inputenc}
\usepackage[T1]{fontenc}
\usepackage{mathptmx}
\usepackage{etoolbox}
\usepackage{mathtools}
\usepackage{scalerel}
\usepackage{braket}

\newcommand{\beq}{\begin{equation}}
\newcommand{\eeq}{\end{equation}}
\makeatletter
\def\@email#1#2{%
 \endgroup
 \patchcmd{\titleblock@produce}
  {\frontmatter@RRAPformat}
  {\frontmatter@RRAPformat{\produce@RRAP{*#1\href{mailto:#2}{#2}}}\frontmatter@RRAPformat}
  {}{}
}%
\makeatother

\bibpunct{}{}{,}{s}{}{\textsuperscript{,}}
\begin{document}

%\preprint{AIP/123-QED}

\title{Employing an operator form of the Rodrigues formula to calculate wavefunctions without differential equations}
% Force line breaks with \\

% \affiliation{Department of Physics, Bowdoin College, Maine USA}%Lines break automatically or can be forced with \\
\author{Joseph R. Noonan}
\affiliation{Department of Physics and Astronomy, Biomedical Physical Sciences 5,
67 Wilson Road, Michigan State University, East Lansing, Michigan 48824}
\author{Maaz ur Rehman Shah}
\affiliation{Department of Computer Sciences, Institute of Business Administration, University Road, Karachi, Sindh 75270, Pakistan}
\author{Luogen Xu}
\author{James. K. Freericks}%
 \email{james.freericks@georgetown.edu}
\affiliation{ 
Department of Physics, Georgetown University, 37th and O Sts. NW, Washington, DC 20057 USA%\\This line break forced with \textbackslash\textbackslash
}%

\date{\today}% It is always \today, today,
             %  but any date may be explicitly specified

\begin{abstract}
The factorization method of Schr\"odinger shows us how to determine the energy eigenstates without needing to determine the wavefunctions in position or momentum space. A strategy to convert the energy eigenstates to wavefunctions is well known for the one-dimensional simple harmonic oscillator by employing the Rodrigues formula for the Hermite polynomials in position or momentum space. In this work, we illustrate how to generalize this approach in a representation-independent fashion to find the wavefunctions of other problems in quantum mechanics that can be solved by the factorization method. We examine three problems in detail: (i) the one-dimensional simple harmonic oscillator; (ii) the three-dimensional isotropic harmonic oscillator; and (iii) the three-dimensional Coulomb problem. This approach can be used in either undergraduate or graduate classes in quantum mechanics.
\end{abstract}

\maketitle
\section{\label{sec:level1}Introduction}

%\color{black}
%Describe the Schroedinger strategy for factorization and the history with supersymmetric quantum mechanics. Explain the importance of determining wavefunctions. Explain how the Rodrigues method is currently being used in quantum textbooks. Explain why its use should be expanded. Summarize the problems we will cover here and why. Summarize the rest of the paper.
%\color{black}

Quantum mechanics is typically taught in one of two approaches---a differential-equation-based approach that uses the Schr\"odinger equation in position space or an algebraic operator-based method which uses abstract operator manipulations to find energy eigenstates. The algebraic method is primarily used for two problems: (i) solving the simple harmonic oscillator in one dimension and (ii) determining states  that have both definite total and $z$-component of angular momentum. In 1940, Schr\"odinger showed how to use algebraic factorization method to solve all exactly solvable quantum problems\cite{schroedinger,schroedinger2,schroedinger3} \color{black}(see Infeld and Hull\cite{infeld-hull} for a review). Schr\"odinger's factorization method was reinvigorated by Witten in his development of supersymmetric quantum mechanics.\cite{witten}~ The factorization method approach, in its simplest form, is what is used in the abstract treatment of the simple harmonic oscillator. Nearly all textbooks that discuss it, will also show how one can find wavefunctions in this approach as well. Most use the subsidiary condition (described in more detail below), given by $\hat{a}|0\rangle=0$, to determine the ground-state wavefunction by converting it into a first-order differential equation in position space. A much smaller subset of quantum textbooks (maybe about 15\%) will also show how the higher-energy eigenstates, given by $(1/\sqrt{n!})\left (\hat{a}^\dagger\right )^n|0\rangle=|n\rangle$ can be expressed as a differential operator (raised to the nth power) acting on the ground-state wavefunction. They then convert the power of operators acting on the ground state, via the Rodrigues formula, into the well-known result for the excited-state wavefunctions in terms of a Hermite polynomial multiplied by a Gaussian. In this work, we show how this approach can be generalized, using an operator-based methodology (as opposed to a differential equation-based methodology), to find the wavefunctions of energy eigenstates in a Rodrigues-formula inspired approach. Note that we can only do this for exactly solvable problems, which are so-called shape-invariant potentials for the operator method. We explicitly cover three problems: (i) the simple harmonic oscillator in one dimension; (ii) the isotropic oscillator in three dimensions; and (iii) the Coulomb problem in three dimensions. We provide shorter summaries for two two-d examples \color{black}in the supplementary material.\cite{supplemental}\color{black}

Here we provide a brief summary of the traditional Rodrigues formulas. There are two of these relevant for this work: (i) the formula for the Hermite polynomials, given by
\begin{equation}
    H_n(x)=(-1)^ne^{x^2}\frac{d^n}{dx^n}\left (e^{-x^2}\right)\label{eq:hermite-rodrigues}
\end{equation}
and (2) the formula for the associated Laguerre polynomials, given by
\begin{equation}
    L_n^{(\alpha)}(x)=\frac{x^{-\alpha}e^x}{n!}\frac{d^n}{dx^n}\left (x^{n+\alpha}e^{-x}\right).\label{eq:laguerre-rodrigues}
\end{equation}
A treatment of the Rodrigues formula from a differential equation point of view is given in Chapter 12 of Arfken, Weber and Harris;\cite{arfken} another way to determine them is by using the Laplace method to solve the confluent hypergeometric equation, where they arise as residues in a contour integral.\cite{canfield}

The remainder of the paper is organized as follows. In Sec. II, we describe in detail how the Schr\"odinger factorization method works and how it can be manipulated to represent a generalized operator form of the Rodrigues formulas for the wavefunctions. In Sec. III, we show how these techniques can be applied to the simple harmonic oscillator in one dimension. Section IV does the same for the isotropic oscillator in three dimensions. Section V covers the Coulomb problem in three dimensions.  In Sec. VI we describe our thoughts on how to present these materials in instruction. We summarize the results in Sec. VII.

\section{Formalism of the factorization method and procedure to relate to Rodrigues formulas\label{factorization method}}

%\color{black}
%Describe how one performs factorization, what the factorization chain is (and the auxiliary Hamiltonians. Derive the intertwining relation. Describe the subsidiary condition and the rules for factorization. Derive the intertwining relation. Verify the energy eigenstate. Normalize the energy eigenstate.

%Discuss how one needs to reduce the product of the string of raising operators into an operator of position. Explain how this can be done via brute force using the subsidiary condition. Explain the alternative practice, which employs the rules for Rodrigues formula: (1) find the similarity transformation to write A as O p O+; (2) find the state that p annihilates; (3) Construct a nested commutator expression for the product of raising operators; (4) determine a recurrence relation for the products; and (5) solve the recurrence relation.
%\color{black}

We briefly review the factorization method of Schr\"odinger\cite{schroedinger} to clarify our notation and to set the stage for how the Rodrigues formula is generalized into an operator form. In the factorization method, we seek to find Hermitian conjugate operators $\hat{A}_0^{\phantom{\dagger}}$ and $\hat{A}_0^\dagger$ such that $\hat{H}=\hat{p}^2/2M+V(\hat{x})=\hat{H}_0=\hat{A}_0^\dagger\hat{A}_0^{\phantom{\dagger}}+E_0$, where we introduce a subscript $0$ to the original Hamiltonian because it will be the first element in the factorization chain. These ladder operators \textit{are not} the same as the conventional ones used for the harmonic oscillator, and in general, their commutator is not equal to one.

The ground-state of $\hat{H}_0$ is the state that satisfies $\hat{A}_0^{\phantom{\dagger}}|\psi_0\rangle=0$, which is called the subsidiary condition. This is the ground state because the operator part of the Hamiltonian is a positive semidefinite operator (all eigenvalues greater or equal to zero) when expressed in terms of the raising and lowering operators. This can be seen by relating expectation values to norms---$\langle \psi|\hat{A}^\dagger\hat{A}|\psi\rangle=|\hat{A}|\psi\rangle|^2\ge 0$ and the only case where it equals zero is if the vector $\hat{A}|\psi\rangle=0$, which is the origin of the subsidiary condition determining the ground state. 
%The state must also be normalizable, and this will be guaranteed by some simple requirements that the ladder operators must satisfy when $|x|\gg 1$. The requirement is best expressed in terms of the superpotential, which is defined below---it must be positive for $x\to\infty$ and negative for $x\to-\infty$. 
To find the excited states, we next form the factorization chain by defining the first auxiliary Hamiltonian $\hat{H}_1=\hat{A}_0^{\phantom{\dagger}}\hat{A}_0^\dagger+E_0=\hat{p}^2/2M+V_1(\hat{x})$, which has the raising and lowering operators reversed. This auxiliary Hamiltonian has a different potential from the original Hamiltonian $V_1(\hat{x})\ne V_0(\hat{x})$, which is determined after explicitly computing $\hat{A}_0^{\phantom{\dagger}}\hat{A}_0^\dagger$, so we factorize it as well, in the form $\hat{H}_1=\hat{A}_1^\dagger\hat{A}_1^{\phantom{\dagger}}+E_1$, where the auxiliary ground state is given by $\hat{A}_1^{\phantom{\dagger}}|\phi_1\rangle=0$. We continue forming new auxiliary Hamiltonians and finding new auxiliary Hamiltonian ground states by repeating this procedure. So, in general, we have that
\begin{align}
&\hat{H}_{i+1}=\frac{\hat{P}^2}{2M}+V_{i+1}(\hat{x})=\hat{A}_i^{\phantom{\dagger}}\hat{A}_i^\dagger+E_i=\hat{A}_{i+1}^\dagger\hat{A}_{i+1}^{\phantom{\dagger}}+E_{i+1}\nonumber\\
&\text{and}~~\hat{A}_{i+1}^{\phantom{\dagger}}|\phi_{i+1}\rangle=0.\label{eq:auxiliary}
\end{align}

The definitions of the factorization chain allow us to construct the intertwining relation, given by
\begin{equation}
    \hat{H}_i\hat{A}_i^\dagger=\hat{A}_i^\dagger\hat{A}_i^{\phantom{\dagger}}\hat{A}_i^\dagger+E_i\hat{A}_i^\dagger=\hat{A}_i^\dagger\hat{H}_{i+1},
\end{equation}
which follows from the definition of the two auxiliary Hamiltonians and factoring the raising operator out to the left or to the right. The intertwining relation allows us to construct the excited states of the original Hamiltonian. Consider the state
\begin{equation}
    |\psi_n\rangle=C_n\hat{A}_0^\dagger\hat{A}_1^\dagger\cdots\hat{A}_{n-1}^\dagger|\phi_n\rangle,\label{eq:norm}
\end{equation}
with $C_n$ a normalization constant that will be determined below. To show that this is an energy eigenstate of $\hat{H}_0$, we simply apply the Hamiltonian to the state from the left. As we move the Hamiltonian to the right through each raising operator, the intertwining relation tells us that the index increases by 1 for each shift, until we get to the end of the product, where we have $\hat{H}_n$ acting on $|\phi_n\rangle$. But that state is the ground state of this auxiliary Hamiltonian, with energy $E_n$. Hence, we learn that the full state is an eigenstate of the original $\hat{H}$ with eigenvalue $E_n$. To find the normalization constant, we simply calculate $\langle\psi_n|\psi_n\rangle$, and replace the innermost $\hat{A}_0^{\phantom{\dagger}}\hat{A}_0^\dagger$ by $\hat{H}_1-E_0$. We then move the Hamiltonian operator to the right using the intertwining relation until it reaches the state on the right, where it is converted to $\hat{H}_n$. It can then act on the state giving $E_n$. Hence, we can remove the factor $\hat{A}_0^{\phantom{\dagger}}\hat{A}_0^\dagger$ and replace it by $E_n-E_0$. We repeat this to remove each pair of lowering-raising operator products and finally determine that
\begin{equation}
    C_n=\frac{1}{\sqrt{(E_n-E_0)(E_n-E_1)\cdots(E_n-E_{n-1})}}. \label{normalization}
\end{equation}

\begin{figure}
    \centering
    \includegraphics[width=0.48\textwidth]{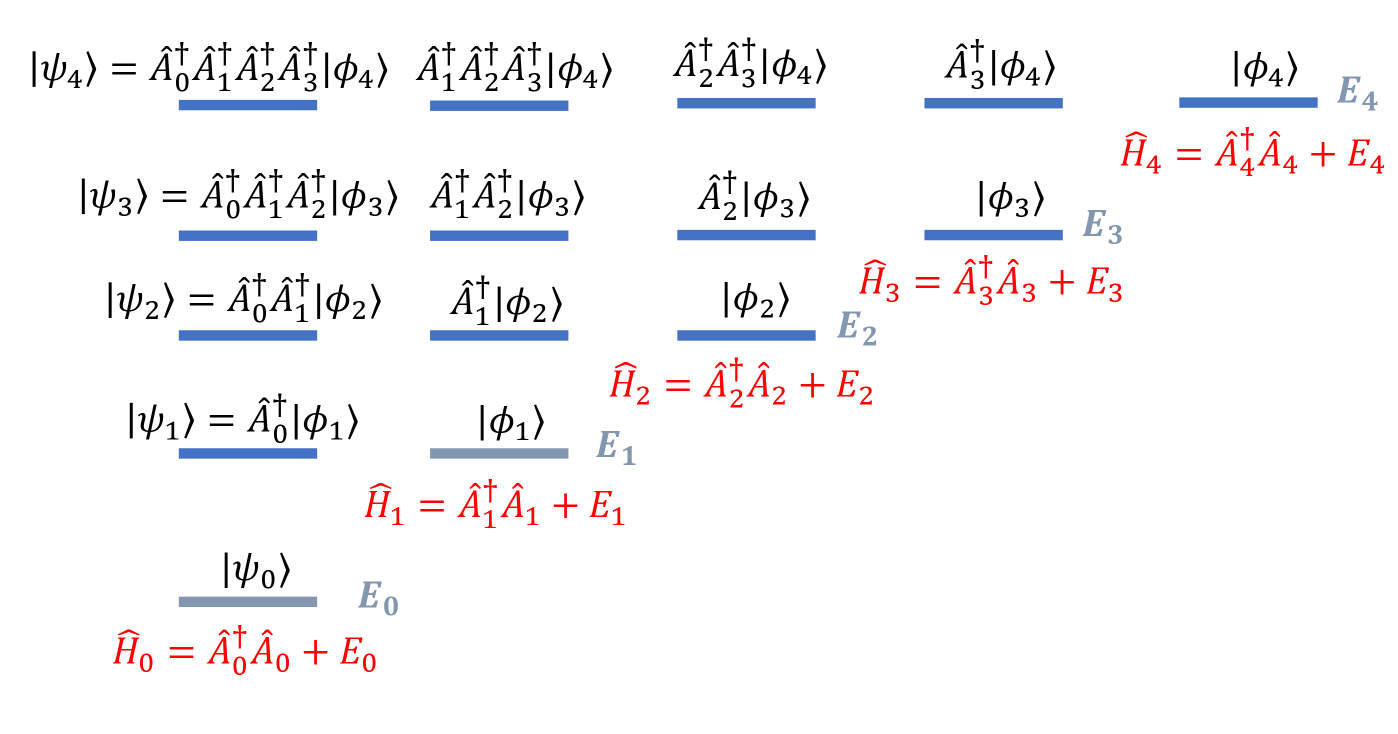}
    \caption{(Color online) Schematic of the factorization chain. On the far left, we have the original Hamiltonian and its ground and excited states as constructed from the factorization by applying strings of raising operators onto auxiliary Hamiltonian ground states as we move horizontally upward from the ground state. As we move to the right, we see the ground and excited states for the first auxiliary Hamiltonian (vertically), then the second and so on. If we instead view the figure along the horizontal lines, we see the different states that are degenerate in energy. For example, the top row shown here starts with the fourth auxiliary Hamiltonian ground state, then the first excited state of the third auxiliary Hamiltonian, and so on until we reach the fourth excited state of the original Hamiltonian. All these states have the same energy $E_4$. In this way, you can see the hidden structure behind every energy eigenvalue problem, where there are other Hamiltonians that share all the bound-state energies except for a finite number of them.\label{fig:chain}}
\end{figure}

This procedure of constructing auxiliary Hamiltonians and finding excited states of the original Hamiltonian via strings of raising operators acting on auxiliary Hamiltonian ground states is called the factorization chain. Each auxiliary Hamiltonian has the same energy eigenvalues as the previous Hamiltonian in the chain, except for the ground state energy eigenvalue. The energies of the states across the chain are all equal to each other, as illustrated in Fig.~\ref{fig:chain}. A summary of the factorization method can be found in Ohanian\cite{ohanian} and Cooper, Khare and Sukhatme.\cite{cooper}
\color{black}

The lowering operator is written in a standard form in terms of the position $\hat{x}$ and momentum $\hat{p}$ operators as 
\begin{equation}
    \hat{A}=\frac{1}{\sqrt{2M}}\big(\hat{p}-i\hbar kW(k'\hat{x})\big),
\end{equation}
where $k$ and $k'$ are real constants with dimensions of inverse length, $M$ is the mass, and the superpotential $W$ is a real-valued function of its dimensionless argument. The name lowering operator does not imply the well-known properties of the lowering operator in the one-d harmonic oscillator. By examining Fig.~\ref{fig:chain} and using the intertwining relation, one can see that applying $\hat{A}_i$ to any eigenstate in the $i$th column, will produce an eigenstate of the same energy in the adjacent column to the right and in the same row. The name superpotential comes from supersymmetric quantum mechanics. We can think of it as an operator valued function and also as a regular function, by replacing $\hat{x}\to x$. The requirement that the auxiliary ground state is normalizable is that $kW(k'x)$ must be positive for $x\to\infty$ and negative for $x\to-\infty$---this condition guarantees the wavefunctions decay faster than any power as $x\to\pm\infty$. We need this condition because the factorization of a given Hamiltonian is not unique, and this condition allows us to determine the correct superpotential to use in each factorization. 

The energy eigenstate is given by the product of a string of raising operators acting on an auxiliary Hamiltonian ground state. The wavefunction in position space is then found by simply multiplying this state by a position bra:
\begin{equation}
    \psi_n(x)=\langle x|\psi_n\rangle.
\end{equation}
We can evaluate this most efficiently if we can convert the product of the string of operators acting on the auxiliary ground state into a set of nested commutators acting on a state derived from the auxiliary Hamiltonian ground state. The strategy to do this is the operator generalization of the Rodrigues formula. Five steps are involved.
\color{black}
\begin{enumerate}
    \item  Rewrite the raising operators as a similarity transformation in the form
\begin{equation}
    \hat{A}_n^\dagger=\frac{1}{\sqrt{2M}}\hat{O}_n\hat{p}\hat{O}_n^{-1},
\end{equation}
with $\hat{O}_n$ some operator that needs to be determined. We will show how to accomplish this in the different examples. Since exactly solvable problems have shape-invariant superpotentials for the ladder operators,\cite{shape-invariant}~ once we solve it for one operator, it is a simple task to solve it for all other ladder operators in the factorization chain. 
\item Find a state, constructed by applying operators onto $|\phi_n\rangle$, that is annihilated by $\hat{p}$. Once we have such a state, we can add $\hat{p}$ times that state to any expression, without changing it, because it corresponds to adding zero. 
\item Use the add-zero property to convert the product of a string of raising operators acting on an auxiliary Hamiltonian ground state into a sequence of nested commutators acting on the state that is annihilated by $\hat{p}$. Since a commutator with $\hat{p}$ acts in the same way as a derivative, the nested commutator object is a form of a Rodrigues formula for a polynomial, namely a derivative raised to the $n$th power acting on a function and then divided by that function. 
\item Determine a recurrence relation for these nested commutators when we compare the sequence for one excited state with the next excited state. 
\item Solve the recurrence relation to determine an expression for the product of raising operators acting on an auxiliary ground state as a function of the position operator acting on the same state. 
\end{enumerate}
\color{black}
This completes the generalization of the Rodrigues formula in terms of operators. 
One can jump from the nested commutators to repeated derivatives, which then become Rodrigues formulas for the different polynomials in the wavefunctions. In this work we carry out all steps in a representation-independent fashion.

This procedure sounds somewhat abstract, so we next describe an explicit example. But note that by following this procedure, we can calculate wavefunctions completely algebraically, without requiring any differential equations. This will also become clearer as we go through the examples.

\section{One-dimensional simple harmonic oscillator}

The simple harmonic oscillator in one dimension is our first example. The Hamiltonian is
\begin{equation}
    \hat{H}=\frac{\hat{p}^2}{2M}+\frac{1}{2}M\omega^2\hat{x}^2,
\end{equation}
with $[\hat{x},\hat{p}]=i\hbar$. Here, $M$ is the mass of the particle and $\omega$ is the frequency of the oscillator. In the Schr\"odinger factorization method, we factorize the Hamiltonian into $\hat{A}^\dagger\hat{A}$, with
\begin{equation}
    \hat{A}=\frac{1}{\sqrt{2M}} (\hat{p}-iM\omega\hat{x})
\end{equation}
and $E=\hbar\omega/2$; the requirement that the superpotential $kW(k'x)=M\omega x/\hbar$ has positive values for $x\to\infty$ and negative values for $x\to-\infty$ (to ensure normalizability) is clearly satisfied. Because $\hat{A}\hat{A}^\dagger=\hat{A}^\dagger\hat{A}+\hbar\omega$, we immediately verify that the auxiliary Hamiltonians in Eq.~(\ref{eq:auxiliary}) have the same functional form for the potential $V_n(\hat{x})$, but each is shifted upward by a constant $n\hbar\omega$ for the $n$th auxiliary Hamiltonian, so that $V_n(\hat{x})=M\omega^2\hat{x}^2/2+n\hbar\omega$. This means the factorization chain produces the same ladder operators for each auxiliary Hamiltonian in the chain; in fact, this is the only Hamiltonian that does this. Hence $\hat{A}_n=\hat{A}$ for all $n$, and the state that satisfies the initial subsidiary condition $\hat{A}|0\rangle=0$ is the auxiliary Hamiltonian ground state for all $n$.

Most students are not familiar with the Schr\"odinger form of the ladder operators, so before proceeding further we convert to the more familiar Dirac form given by
\begin{equation}
    \hat{a}=\frac{i}{\sqrt{\hbar\omega}}\hat{A}=\sqrt{\frac{M\omega}{2\hbar}}\left (\hat{x}+i\frac{\hat{p}}{M\omega}\right ).
\end{equation}
Using this Dirac form, we have the familiar results: (i) the subsidiary condition is $\hat{a}|0\rangle=0$; (ii) the $n$th excited state is $|n\rangle=(1/\sqrt{n!})\left (\hat{a}^\dagger\right )^n|0\rangle$; and (iii) the $n$th energy eigenvalue is $E_n=\hbar\omega\left (n+1/2\right )$.

Now we work out the procedure to generalize the Rodrigues formula to an operator format. First we need to find the appropriate similarity transformation. One can use the Hadamard lemma (for example, see Ch. 3 of Merzbacher\cite{hadamard})\color{black}
\begin{equation}
    e^{\hat{A}}\hat{B}e^{-\hat{A}}=\hat{B}+[\hat{A},\hat{B}]+\tfrac{1}{2}[\hat{A},[\hat{A},\hat{B}]]+\tfrac{1}{3!}[\hat{A},[\hat{A},[\hat{A},\hat{B}]]]+\cdots
\end{equation}
which relates the similarity transformation of $\hat{B}$ to an infinite series of increasingly nested commutators. We use it to determine the similarity transformation of $\hat{p}$ that produces $\hat{a}^\dagger$. After a little trial and error, we find that
\begin{equation}
    \hat{a}^\dagger=-\tfrac{i}{\sqrt{2\hbar M\omega}}e^{\frac{M\omega}{2\hbar}\hat{x}^2}\hat{p}e^{-\frac{M\omega}{2\hbar}\hat{x}^2}.
\end{equation}
\color{black}
The Hadamard lemma truncates after two terms here, yielding the lowering operator. 
Next, we find the state annihilated by $\hat{p}$ by taking the Hermitian conjugate of the above similarity transformation, given by
\begin{equation}
    \hat{a}=\tfrac{i}{\sqrt{2\hbar M\omega}}e^{-\frac{M\omega}{2\hbar}\hat{x}^2}\hat{p}e^{\frac{M\omega}{2\hbar}\hat{x}^2},\label{eq:sim}
\end{equation}
and use it in the subsidiary condition ($\hat{a}|0\rangle=0$), multiplied from the left by the appropriate operator. This gives us
\begin{equation}
    -i\sqrt{2\hbar M\omega}\,e^{\frac{M\omega}{2\hbar}\hat{x}^2}\hat{a}|0\rangle=0=\hat{p}\underbrace{e^{\frac{M\omega}{2\hbar}\hat{x}^2}|0\rangle}_{\text{state annihilated by }\hat{p}}.
\end{equation}
Now, we work on the $n$th excited state using the series of steps that are explained below,
\begin{align}
    |n\rangle&=\tfrac{1}{\sqrt{n!}}\tfrac{(-i)^n}{(2\hbar M\omega)^{n/2}}e^{\frac{M\omega}{2\hbar}\hat{x}^2}\hat{p}^ne^{-\frac{M\omega}{2\hbar}\hat{x}^2}|0\rangle\nonumber\\
    &=\tfrac{1}{\sqrt{n!}}\tfrac{(-i)^n}{(2\hbar M\omega)^{n/2}}e^{\frac{M\omega}{2\hbar}\hat{x}^2}\hat{p}^ne^{-\frac{M\omega}{\hbar}\hat{x}^2}e^{\frac{M\omega}{2\hbar}\hat{x}^2}|0\rangle\nonumber\\
    &=\tfrac{1}{\sqrt{n!}}\tfrac{(-i)^n}{(2\hbar M\omega)^{n/2}}e^{\frac{M\omega}{2\hbar}\hat{x}^2}\nonumber\\
    &~~~~~\times\left [\hat{p},\left [\hat{p},\cdots,\left [\hat{p},e^{-\frac{M\omega}{\hbar}\hat{x}^2}\right ]\cdots\right ]\right ]_ne^{\frac{M\omega}{2\hbar}\hat{x}^2}|0\rangle\nonumber\\
    &=\tfrac{1}{\sqrt{n!}}\tfrac{(-i)^n}{(2\hbar M\omega)^{n/2}}e^{\frac{M\omega}{\hbar}\hat{x}^2}\nonumber\\
    &~~~~~\times\left [\hat{p},\left [\hat{p},\cdots,\left [\hat{p},e^{-\frac{M\omega}{\hbar}\hat{x}^2}\right ]\cdots\right ]\right ]_n|0\rangle.\label{eq:nested-sho})
\end{align}
In the first line, we substitute in for $\hat{a}^\dagger$ using the similarity transformation in Eq.~(\ref{eq:sim}) and cancel ``interior'' Gaussian operator factors that multiply to one. In the second line, we use a ``multiply-by-one'' to introduce the state annihilated by $\hat{p}$. In the third line, we start from the rightmost momentum operator and replace 
\begin{equation}
    \hat{p}e^{-\frac{M\omega}{\hbar}\hat{x}^2}e^{\frac{M\omega}{2\hbar}\hat{x}^2}|0\rangle= [\hat{p},e^{-\frac{M\omega}{\hbar}\hat{x}^2} ]e^{\frac{M\omega}{2\hbar}\hat{x}^2}|0\rangle,
\end{equation} 
because this is the same as adding zero due to the fact that $\hat{p}$ annihilates the state $\exp(M\omega\hat{x}^2/2\hbar)|0\rangle$. We then repeat this procedure with the next momentum operator $n-1$ more times to obtain the $n$-fold nested commutator. Finally, in the fourth line, we combine the operator factors that depend on $\hat{x}$, because the nested commutator is a function of $\hat{x}$ only so it commutes with $\hat{x}$. Recall, the commutator of a function of $\hat{x}$ with $\hat{p}$ produces a derivative of the function of $\hat{x}$ multiplied by numbers, so it is a function of $\hat{x}$; this holds for nested commutators with momentum too.

Readers familiar with Rodrigues formulas will already recognize that this result looks similar to the Rodrigues formula for Hermite polynomials.  At this stage, one can directly get to the Rodrigues formula in differential form by replacing $\hat{p}\to -i\hbar d/dx$ and noting that nested commutators become multiple derivatives. Then one can use the Rodrigues formula in Eq.~(\ref{eq:hermite-rodrigues}) to complete the derivation. However, we want to establish it without derivatives, using just operators. \color{black} So, we define a polynomial $H_n$ in terms of the nested commutators in Eq.~(\ref{eq:nested-sho}) and show that it is a Hermite polynomial by verifying its recurrence relation. This requires introducing some constants to agree with the standard definitions. We define
\begin{align}
    H_n\left (\hat{x}\sqrt{\tfrac{M\omega}{\hbar}}\right )|0\rangle&=\frac{(-i)^n}{(\hbar M\omega)^{n/2}}e^{\frac{M\omega}{\hbar}\hat{x}^2}\nonumber\\
    &\times\left [\hat{p},\left [\hat{p},\cdots,\left [\hat{p},e^{-\frac{M\omega}{\hbar}\hat{x}^2}\right ]\cdots\right ]\right ]_n|0\rangle\nonumber\\
    &=\sqrt{2^n}\left (\hat{a}^\dagger\right )^n|0\rangle=\sqrt{2^nn!}|n\rangle.\label{H_n}
\end{align}
Setting $n=0$, we find $H_0(\hat{x}\sqrt{M\omega/\hbar})=1$. Similarly, setting $n=1$, and computing $  [ \hat{p},\exp(-M\omega\hat{x}^2/\hbar)]=2iM\omega\hat{x}\exp(-M\omega\hat{x}^2/\hbar)$, which can be most easily worked out using the Hadamard lemma, gives us $H_1(\hat{x}\sqrt{M\omega/\hbar})=2\hat{x}\sqrt{M\omega/\hbar}$. These are the first two Hermite polynomials (with the so-called physicist normalization). The general recurrence relation is found using the series of steps that are explained below,
\begin{align}
    H_{n+1}\left (\hat{x}\sqrt{\tfrac{M\omega}{\hbar}}\right )\ket0 &= \tfrac{-i}{\sqrt{\hbar M\omega}}e^{\frac{M\omega}{\hbar}\hat{x}^2}\nonumber\\
    &\times \left[\hat p, e^{-\frac{M\omega}{\hbar}\hat{x}^2} H_n\left (\hat{x}\sqrt{\tfrac{M\omega}{\hbar}}\right )\right]\ket0 \nonumber\\
    &= 2\hat{x}\sqrt{\tfrac{M\omega}{\hbar}}H_{n}\left (\hat{x}\sqrt{\tfrac{M\omega}{\hbar}}\right )\ket0 \nonumber\\
    &- \tfrac{i}{\sqrt{\hbar M\omega}}\left[\hat{p},H_{n}\left (\hat{x}\sqrt{\tfrac{M\omega}{\hbar}}\right )\right]\ket0
\end{align}
The first line comes directly from the definition in Eq.~(\ref{H_n}); one needs to use manipulations that move the exponential operator $\exp(M\omega\hat{x}^2/2\hbar)$ to the right and use the properties of the state annihilated by $\hat{p}$ to establish this result and the second comes from applying the Leibniz rule for products of operators in a commutator given by $[\hat{A},\hat{B}\hat{C}]=\hat{B}[\hat{A},\hat{C}]+[\hat{A},\hat{B}]\hat{C}$. Using the fact that $\hat{p}=(\hat{a}-\hat{a}^\dagger)\sqrt{\hbar M\omega/2}/i=[2\hat{a}-(\hat{a}+\hat{a}^\dagger)]\sqrt{\hbar M\omega/2}/i$ as well as the fact that $\hat{a}+\hat{a}^\dagger$ commutes with $\hat{x}$, we can convert the last term into
\begin{align}
    -\sqrt2\left[\hat{a},H_{n}\left (\hat{x}\sqrt{\tfrac{M\omega}{\hbar}}\right )\right]\ket0 &=-\sqrt2\hat{a}H_{n}\left (\hat{x}\sqrt{\tfrac{M\omega}{\hbar}}\right )\ket0\nonumber\\
    &= -\sqrt{2^{n+1}}\hat{a}(\hat{a}^\dagger)^n\ket0\nonumber\\
    & = -\sqrt{2^{n+1}}n(\hat{a}^\dagger)^{n-1}\ket0.
\end{align}
The first line uses the subsidiary condition to convert the commutator into just the first term of the commutator, the second line applies Eq.~(\ref{H_n}), and the third line uses the fact that $[\hat{a},(\hat{a}^\dagger)^n]=n(\hat{a}^\dagger)^{n-1}$ and employs the subsidiary condition for an ``add-zero'' again. Thus, we have that
\begin{align}
    H_{n+1}\left (\hat{x}\sqrt{\tfrac{M\omega}{\hbar}}\right )\ket0 
    &= 2\hat{x}\sqrt{\tfrac{M\omega}{\hbar}}H_{n}\left (\hat{x}\sqrt{\tfrac{M\omega}{\hbar}}\right )\ket0 \nonumber\\
    &- 2nH_{n-1}\left (\hat{x}\sqrt{\tfrac{M\omega}{\hbar}}\right )\ket0,
\end{align}
which is the same as the recurrence relation for Hermite polynomials. Using induction then implies that $H_n (\hat{x}\sqrt{M\omega/\hbar} )|0\rangle$ is indeed equal to an operator-valued Hermite polynomial acting on $\ket0.$ The wavefunction is found by multiplying with a position bra from the left, so 
\begin{equation}
  \Psi_n(x)\propto\langle x|H_n\left (\hat{x}\sqrt{\tfrac{M\omega}{\hbar}}\right )|0\rangle=H_n\left (x\sqrt{\tfrac{M\omega}{\hbar}}\right )\langle x|0\rangle.
\end{equation}
To finish the calculation, we need to determine $\langle x|0\rangle$.\par
The ground state wavefunction is found by using a multiply-by-one with two Gaussian operators, moving one out of the matrix element by evaluating it against the position bra, using the translation operator to write $\langle x|$ as the translation of the position eigenstate at the origin $\langle 0_x|$ to the state $\langle x|$, recognizing that the translation operator can be replaced by unity when acting on the state to its right (because $\hat{p}$ annihilates that state), and finally acting the Gaussian operator against $\langle 0_x|$ where it is replaced by 1. Hence,
\begin{align}
    &\langle x|0\rangle = \bra{x}e^{-\frac{M\omega}{2\hbar}\hat{x}^2}e^{\frac{M\omega}{2\hbar}\hat{x}^2}\ket0=e^{-\frac{M\omega}{2\hbar}x^2}\bra{x}e^{\frac{M\omega}{2\hbar}\hat{x}^2}\ket0\nonumber\\
    &~~= e^{-\frac{M\omega}{2\hbar}x^2}\bra{0_x}e^{\frac{i}{\hbar}x\hat{p}}e^{\frac{M\omega}{2\hbar}\hat{x}^2}\ket0 = e^{-\frac{M\omega}{2\hbar}x^2}\bra{0_x}e^{\frac{M\omega}{2\hbar}\hat{x}^2}\ket0\nonumber\\
    &~~= e^{-\frac{M\omega}{2\hbar}x^2}\langle0_x|0\rangle.
\end{align}
By normalizing, we find that $\langle0_x|0\rangle=\sqrt[4]{M\omega/\pi\hbar}$, so the final result with all constant factors included is 
\begin{equation}
    \psi_n(x) = \tfrac1{\sqrt{n!2^n}}\left(\tfrac{M\omega}{\pi\hbar}\right)^{1/4}H_{n}\left(x\sqrt{\tfrac{M\omega}{\hbar}}\right)e^{-\frac{M\omega}{2\hbar}x^2},
\end{equation}
which is the well-known result.

One can also perform this derivation in momentum space, which would be a good exercise to assign to students. In general, the similarity transformation cannot be performed on the ladder operators in terms of a similarity transformation with respect to $\hat{x}$, but the harmonic oscillator is the one case where this can be done.

\section{Three-dimensional isotropic oscillator}

The simplest way to solve the three-dimensional isotropic harmonic oscillator using the factorization method is to use separation of variables to split the Hamiltonian into its radial and angular components: 
\begin{align}
    \hat H = \frac{\hat p^2}{2M} + \frac12M\omega\hat r^2 = \frac{\hat p_r^2}{2M} + \frac{\hat L^2}{2M\hat r^2} + \frac12M\omega\hat r^2.
\end{align}
We label the angular momentum eigenstates as $\ket{l,m}$ and define
\begin{align}
    \hat H_l = \frac{\hat p_r^2}{2M} + \frac{\hbar^2 l(l+1)}{2M\hat r^2} + \frac12M{\omega}^2\hat r^2,
\end{align}
so that $\hat H(\ket\psi\otimes\ket{l,m})=(\hat H_l\ket{\psi})\otimes\ket{l,m}$, which is just separation of variables in Dirac form.\cite{spherical-translation-operator}~ Here $\hat{p}_r$ is the radial momentum operator given by $(1/\hat{r})\hat{\vec{r}}\cdot\hat{\vec{p}}-i\hbar/\hat{r}$. $\hat H_l$ can be factored with the lowering operator
\begin{align}
    \hat A_l = \frac1{\sqrt{2M}}\left(\hat p_r +\frac{i\hbar(l+1)}{\hat r} - iM\omega\hat r\right),
\end{align}
with energy $E_l=\hbar\omega(l+3/2)$; for assistance with working out commutators, please see the appendix of Ref.~\onlinecite{spherical-translation-operator}. In this case, we find $\hat{A}_l\hat{A}_l^\dagger+E_l=\hat{H}_{l+1}+\hbar\omega$, so the $k$th auxiliary Hamiltonian for $\hat{H}_l$ is just $\hat{H}_{l+k}+k\hbar\omega$. Thus the eigenstates of $\hat{H}_l$ are given by $C_{k,l}\hat{A}^\dagger_l\cdots\hat{A}^\dagger_{l+k-1}\ket{\phi_{l+k}}$, where $\ket{\phi_l}$ represents the ground state of $\hat{H}_l$. Note that one needs to be very careful to keep straight the difference between the Hamiltonians for definite angular momentum versus the auxiliary Hamiltonians for each problem with definite angular momentum; they are related to each other by constant shifts proportional to $\hbar\omega$.\par
Applying the Hadamard lemma as well as the fact that $\hat{r}^k\hat{p}_r\hat{r}^{-k}=\hat p_r +i\hbar k/\hat{r}$, we can re-express the raising operator as
\begin{align}
    \hat{A}_l^\dagger = \frac1{\sqrt{2M}}e^{\frac{M\omega}{2\hbar}\hat{r}^2}\frac1{\hat r^{l+1}}\hat{p}_r\hat{r}^{l+1}e^{-\frac{M\omega}{2\hbar}\hat{r}^2}.
\end{align}
One can employ the commutator identities in the appendix of Ref.~\onlinecite{spherical-translation-operator} to compute these commutators without converting to differential operators.
We likewise convert $\hat{A}_l$ into \begin{equation}
\frac1{\sqrt{2M}}e^{-\frac{M\omega}{2\hbar}\hat{r}^2}{\hat r^{l}}\left(\hat{p}_r+\frac{i\hbar}{\hat r}\right)\frac1{\hat r^{l}}e^{\frac{M\omega}{2\hbar}\hat{r}^2},
\end{equation}
giving us the subsidiary condition
\begin{align}
    \left(\hat{p}_r+\frac{i\hbar}{\hat r}\right)\frac1{\hat r^l}e^{\frac{M\omega}{2\hbar}\hat{r}^2}\ket{\phi_l}=0.
\end{align}
The reason we write this with a specific shift of the radial momentum operator will become clear in the course of the derivation.

Now, using our formula for the $k$th eigenstate $\ket{\psi_{k,l}}$ of $\hat H_l$, we find that
\begin{align}
    \ket{\psi_{k,l}} = \frac{C_{k,l}}{\sqrt{2M}^k} e^{\frac{M\omega}{2\hbar}\hat{r}^2}\frac1{\hat r^{l+1}}\left(\hat{p}_r\tfrac1{\hat r}\right)^{k}\hat{r}^{l+k+1}e^{-\frac{M\omega}{2\hbar}\hat{r}^2}\ket{\phi_{l+k}},
\end{align}
after canceling out the factors of $\exp(\pm M\omega\hat{r}^2/2\hbar)$ and opposite powers of $\hat r$ from adjacent raising operators. If we look only at the last factor of $\hat{p}_r/\hat r$ and everything that comes after it, we find
\begin{align}
    &\hat{p}_r\hat{r}^{l+k}e^{-\frac{M\omega}{2\hbar}\hat{r}^2}\ket{\phi_{l+k}} = \tfrac1{\hat r}\left(\hat p_r+\tfrac{i\hbar}{\hat r}\right)\hat{r}^{l+k+1}e^{-\frac{M\omega}{2\hbar}\hat{r}^2}\ket{\phi_{l+k}}\nonumber\\
    &~~~~= \left[\tfrac1{\hat r}\left(\hat p_r+\tfrac{i\hbar}{\hat r}\right),e^{-\frac{M\omega}{\hbar}\hat{r}^2}\hat{r}^{2l+2k+1}\right]
    \hat{r}^{-l-k}e^{\frac{M\omega}{2\hbar}\hat{r}^2}\ket{\phi_{l+k}} \nonumber\\
    &~~~~= \left[\tfrac1{\hat r}\hat{p}_r,e^{-\frac{M\omega}{\hbar}\hat{r}^2}\hat{r}^{2l+2k+1}\right]\hat{r}^{-l-k}e^{\frac{M\omega}{2\hbar}\hat{r}^2}\ket{\phi_{l+k}}.
\end{align}
In the first line, we move one factor of $1/\hat{r}$ to the left, shifting the radial momentum by its commutator with $1/\hat{r}$. In the second line, we use a multiply-by-one to create the state annihilated by $\hat{p}_r+i\hbar/\hat{r}$, so we can introduce the commutator. In the third line we recognize that functions of $\hat{r}$ commute with $1/\hat{r}$.
Repeating this with each additional factor of $\hat{p}_r/\hat r$ gives us the operator form of the Rodrigues formula
\begin{align}
    &~~~~\ket{\psi_{k,l}} = \frac{C_{k,l}}{\sqrt{2M}^k} e^{\frac{M\omega}{2\hbar}\hat{r}^2}\frac1{\hat r^{l+1}}\nonumber\\
    &~~~~\times\left[\tfrac1{\hat r}\hat{p}_r,...\left[\tfrac1{\hat r}\hat{p}_r,e^{-\frac{M\omega}{\hbar}\hat{r}^2}\hat{r}^{2l+2k+1}\right]\cdots\right]_k\nonumber\\
    &~~~~\times\hat{r}^{-l-k}e^{\frac{M\omega}{2\hbar}\hat{r}^2}\ket{\phi_{l+k}}\nonumber\\
    &~~~~= \frac1{\hat r^k}\left(\frac{-2i\hbar}{\sqrt{2M}}\right)^k e^{\tfrac{M\omega}{\hbar}\hat{r}^2} \left (\tfrac{M\omega}{\hbar}\hat{r}^2\right )^{-l-\frac12}\nonumber\\
    \times&\left[\tfrac1{\hat r}\tfrac{i\hat{p}_r}{2M\omega},...\left[\tfrac1{\hat r}\tfrac{i\hat{p}_r}{2M\omega},e^{-\tfrac{M\omega}{\hbar}\hat{r}^2}\left(\tfrac{M\omega}{\hbar}\hat{r}^2\right )^{k+l+\frac12}\right]\cdots\right]_k |\phi_{l+k}\rangle.
\end{align}
If one wants to work with the differential form of the Rodrigues formula, one can again convert the nested commutators into multiple derivatives and employ Eq.~(\ref{eq:laguerre-rodrigues}) to determine the associated Laguerre polynomials in the solution.
Instead, we  introduce
\begin{align}
    &L_k^{(l+\frac{1}{2})}\left (\tfrac{M\omega}{\hbar}\hat{r}^2\right )= \frac{\left (\frac{M\omega}{\hbar}\hat{r}^2\right )^{-l-\frac{1}{2}}e^{\frac{M\omega}{\hbar}\hat{r}^2}}{k!}\nonumber\\
    &~~\times\left[\tfrac1{\hat r}\tfrac{i\hat{p}_r}{2M\omega},\cdots\left[\tfrac1{\hat r}\tfrac{i\hat{p}_r}{2M\omega},e^{-\frac{M\omega}{\hbar}\hat{r}^2}\left (\tfrac{M\omega}{\hbar}\hat{r}^2\right)^{k+l+\frac{1}{2}}\right]\cdots\right]_k.\label{Laguerre}
\end{align}
We would like to show that $L_k^{(l+1/2)}$ is equal to an associated Laguerre polynomial, which we will once again do by a recurrence relation. To do this, we need to show that $L_0^{(l+1/2)}(x)=1$ and verify the general recurrence relation
\begin{equation}
    kL_k^{(l+\frac{1}{2})}(x)=(k+l+\tfrac{1}{2})L_{k-1}^{(l+\frac{1}{2})}(x)-xL_{k-1}^{(l+\frac{3}{2})}(x) \label{recurrence 1}
\end{equation}
for Laguerre polynomials. Starting with $k=0$, we see that there is no commutator but the power of $\hat{r}$ and the Gaussian factor remain, so one can see all factors cancel out and we indeed find the Laguerre polynomial is 1. If we evaluate the innermost commutator in Eq.~(\ref{Laguerre}) by using Leibniz's product rule, we find 
\begin{align}
&\left[\tfrac1{\hat r}\tfrac{i\hat{p}_r}{2M\omega},e^{-\frac{M\omega}{\hbar}\hat{r}^2}\left (\tfrac{M\omega}{\hbar}\hat{r}^2\right)^{k+l+\frac{1}{2}}\right]=\nonumber\\
    &~~~~~~~~~\frac{ie^{-\frac{M\omega}{\hbar}\hat{r}^2}}{2M\omega\hat r}\left(\tfrac{M\omega}{\hbar}\right)^{k+l+\frac{1}{2}}\left [\hat{p}_r,\hat{r}^{2k+2l+1}\right ]\nonumber\\
    &~~~~~~+\frac{i}{2M\omega\hat r}\left[\hat{p}_r,e^{-\frac{M\omega}{\hbar}\hat{r}^2}\right]\left (\tfrac{M\omega}{\hbar}\hat{r}^2\right )^{k+l+\frac{1}{2}}\nonumber\\
    &~~~~~~= e^{-\frac{M\omega}{\hbar}\hat{r}^2}(k+l+\tfrac{1}{2})\left (\tfrac{M\omega}{\hbar}\hat{r}^2\right )^{k+l-\frac{1}{2}}\nonumber\\
    &~~~~~~-e^{-\frac{M\omega}{\hbar}\hat{r}^2}\left (\tfrac{M\omega}{\hbar}\hat{r}^2\right )^{k+l+\frac{1}{2}}.
\end{align}
Thus, multiplying Eq.~(\ref{Laguerre}) by $k$ and using it again to define the Laguerre polynomials with different indices, we find that
\begin{align}
    kL_k^{(l+\frac{1}{2})}\left (\tfrac{M\omega}{\hbar}\hat{r}^2\right )&= (k+l+\tfrac{1}{2})L_{k-1}^{(l+\frac{1}{2})}\left (\tfrac{M\omega}{\hbar}\hat{r}^2\right )\nonumber\\
    &-\left (\tfrac{M\omega}{\hbar}\hat{r}^2\right ) L_{k-1}^{(l+\frac{3}{2})}\left (\tfrac{M\omega}{\hbar}\hat{r}^2\right ).
\end{align}
Hence, the operator definition of the Laguerre polynomials does properly satisfy the recurrence relation of the Laguerre polynomials. This then allows us to replace the iterated commutator by the Laguerre polynomial via
\begin{align}
    \ket{\psi_{k,l}}= C_{k,l} \left(-\frac{2i\hbar}{\sqrt{2M}}\right)^k\frac{k!}{\hat{r}^k}L_k^{(l+\frac12)}\left (\tfrac{M\omega}{\hbar}\hat{r}^2\right )\ket{\phi_{l+k}}.
\end{align}
The normalization constant $C_{k,l}$ is found from Eq.~(\ref{eq:norm}); one must be careful to use the auxiliary Hamiltonians in this calculation and recall that they are shifted upwards by constants proportional to integers times $\hbar\omega$. We then find that
\begin{align}
    C_{k,l}=\frac1{\sqrt{(2\hbar\omega)^kk!}}.
\end{align}

What remains is to determine the wavefunction for the ground state. Following a similar strategy as before, we have that
\begin{align}
    \langle{r}|\phi_l\rangle &= r^le^{-\frac{M\omega}{2\hbar}\hat{r}^2}\bra{r}\hat{r}^{-l}e^{\frac{M\omega}{2\hbar}\hat{r}^2}\ket{\phi_l} \nonumber\\
    &= r^le^{-\frac{M\omega}{2\hbar}{r}^2}\bra{0_r}e^{\frac{i}{\hbar}r(\hat{p}_r+\frac{i\hbar}{\hat r})}\hat{r}^{-l}e^{\frac{M\omega}{2\hbar}\hat{r}^2}\ket{\phi_l} \nonumber\\
    &= r^le^{-\frac{M\omega}{2\hbar}{r}^2}\bra{0_r}\hat{r}^{-l}\ket{\phi_l}
\end{align}
The second line replaces $\bra{r}$ by the $r=0$ bra, $\bra{0_r}$ times an exponential of the spherical translation operator,\cite{spherical-translation-operator}
 and the third line uses the fact that the state to the right is annihilated by $\hat{p}_r+i\hbar/\hat{r}$, which replaces the spherical translation operator by 1 and then operates $\exp(M\omega\hat{r}^2/2\hbar)$ to the left onto $\bra{0_r}$ where it is replaced by 1 as well. The remaining term, $\bra{0_r}\hat{r}^{-l}\ket{\phi_l}$ is a constant, which we can calculate by normalizing the wavefunction, giving
\begin{align}
    \bra{0_r}\hat{r}^{-l}\ket{\phi_l} = \left(\frac{M\omega}{\hbar}\right)^{\frac{l}2+\frac34}\frac{2^{\frac{l}2+1}}{\pi^{\frac14}\sqrt{(2l+1)!!}}.
\end{align}
One might have thought that this term diverges, but because the wavefunction behaves as $r^l$ for $r\to 0$, it is the well-defined coefficient of this term.
Summarizing, our work gives us the radial wave function
\begin{align}
    \psi_{k,l}(r) &=
    \tfrac1{\pi^{\frac14}}\sqrt{\tfrac{2^{l+k+2}k!}{(2(l+k)+1)!!}}\left(\tfrac{M\omega}{\hbar}\right)^{\frac{l}2+\frac34} \nonumber\\
    &\times  L_k^{(l+\frac12)}\left(\tfrac{M\omega}{\hbar}r^2\right) r^l e^{-\frac{M\omega}{2\hbar}r^2},
\end{align}
where we have omitted the overall phase factor of $(-i)^k$. The wave function is more commonly expressed in terms of the principle quantum number. Since the $k$th auxiliary Hamiltonian for $\hat{H}_l$ is $\hat{H}_{l+k}+k\hbar\omega$, the energy of $\ket{\psi_{k,l}}$ is the ground state energy of $\hat{H}_{l+k}+k\hbar\omega$, which is $(l+2k+3/2)\hbar\omega$. Hence, we introduce the principle quantum number $n=l+2k$ to obtain our final result
\begin{align}
    \psi_{n,l}(r) &= \frac1{\pi^{\frac14}}\sqrt{\frac{2^{\frac{n+l}2+2}(\frac{n-l}2)!}{(n+l+1)!!}}\left(\tfrac{M\omega}{\hbar}\right)^{\frac{l}2+\frac34} \nonumber\\
    &\times  L_{\frac{n-l}2}^{(l+\frac12)}\left(\tfrac{M\omega}{\hbar}r^2\right) r^l e^{-\frac{M\omega}{2\hbar}r^2},
\end{align}
which is now in standard form for the radial wavefunction.

The problem can also be solved in momentum space, but this might be too challenging for students to carry out. This is not because it is too difficult, but rather because it might be unfamiliar to them. This is again a case where the similarity transformation can be made with respect to the $\hat{r}$ function for the ladder operators, which allows for a very similar solution procedure. But determining the proper definition for the Laguerre polynomials from the nested commutators and solving the recurrence relation would be a challenge for students unless properly scaffolded.

\section{Coulomb problem for hydrogen}

Like the 3D harmonic oscillator problem, it is easiest to solve the Coulomb problem by using the rotational symmetry and separation of variables, so that
\begin{equation}
    \hat H = \frac{\hat p^2}{2M} - \frac{e^2}{\hat r} = \frac{\hat p_r^2}{2M} + \frac{\hat L^2}{2M\hat r^2} - \frac{e^2}{\hat r}.
\end{equation}
We can define the Hamiltonians with constant angular momentum as
\begin{equation}
    \hat H_l = \frac{\hat p_r^2}{2M} + \frac{\hbar^2l(l+1)}{2M\hat r^2} - \frac{e^2}{\hat r},
\end{equation}
so that once again $\hat H(|\psi\rangle\otimes|l,m\rangle)=(\hat H_l|\psi\rangle)\otimes|l,m\rangle$. We can factor $\hat H_l$ with the lowering operator given by
\begin{equation}
    \hat A_l = \frac1{\sqrt{2M}}\left(\hat p_r+\frac{i\hbar(l+1)}{\hat r}-\frac{i\hbar}{(l+1)a_0}\right),
\end{equation}
and energy $E_{l,0}=-e^2/[2(l+1)^2a_0]$. This time, $\hat A_l^{\phantom{\dagger}}\hat A_l^\dagger+E_{l,0}=\hat H_{l+1}$, so the auxiliary Hamiltonian for $\hat H_l$ is $\hat H_{l+1}$. Therefore, the $k$th excited state with angular momentum $l$ is $C_{l,k}\hat A_l^\dagger\cdots\hat A_{l+k-1}^\dagger\ket{\phi_{l+k}}$ (where $\ket{\phi_{l+k}}$ is the ground state of $\hat H_{l+k}$ and has energy $E_{l,k}=-e^2/[2(l+k+1)^2a_0]$. However, we usually define these states not in terms of $l$ and $k$, but in terms of the principal quantum number $n$, so that $E_n=-e^2/[2n^2a_0]$. The state $\ket{n,l}$ corresponds to $k=n-l-1$. Thus, we write 
\begin{equation}
    \ket{n,l}=C_{n,l}\hat A_l^\dagger\cdots\hat A_{n-2}^\dagger\ket{n,n-1}.
\end{equation}
Using the Hadamard lemma, we re-express the raising operator as a similarity transformation via
\begin{equation}
    \hat A_l^\dagger = \frac1{\sqrt{2M}}\frac1{\hat r^{l+1}}e^{\frac{\hat r}{(l+1)a_0}}\hat p_re^{-\frac{\hat r}{(l+1)a_0}}\hat r^{l+1},
\end{equation}
and the subsidiary condition becomes
\begin{equation}
    \hat p_r e^{\frac{\hat r}{na_0}}\frac1{\hat r^n}\ket{n,n-1}=0.
\end{equation}

Therefore,
\begin{align}
    \ket{n,l} &= \frac{C_{n,l}}{\sqrt{2M}^{n-l-1}} \frac1{\hat r^{l+1}}e^{\frac{\hat r}{(l+1)a_0}}\nonumber\\
    &\times \prod_{j=1}^{n-l-1}\left(\hat p_r e^{-\frac{\hat r}{a_0}\left(\frac1{l+j}-\frac1{l+j+1}\right)}\frac1{\hat r}\right)e^{-\frac{\hat r}{na_0}}\hat r^n\ket{n,n-1}.
\end{align}
The product contains terms when $l\ne n-1$. We take the rightmost factor of the product and act it on everything to the right, giving us the commutator replacement
\begin{align}
    &\hat p_r e^{-\frac{\hat r}{(n-1)a_0}}\hat r^{n-1}\ket{n,n-1}\nonumber\\
    =& \left[\hat p_r, e^{-\frac{\hat r}{a_0}\frac{2n-1}{n(n-1)}}\hat r^{2n-1}\right]e^{\frac{\hat r}{na_0}}\frac1{\hat r^n}\ket{n,n-1},
\end{align}
after recalling the state annihilated by $\hat{p}_r$. We continue pulling out factors from the rightmost terms in the remaining products to similarly convert to additional nested commutators. We obtain $|n,l\rangle=\hat{M}_{n,l}|n,n-1\rangle$, with 
\begin{align}
    \hat{M}_{n,l} &= \frac{C_{n,l}}{\sqrt{2M}^{n-l-1}}\frac1{\hat r^{l+1}}e^{\frac{\hat r}{(l+1)a_0}}
    \left[\hat p_r, e^{-\frac{\hat r}{a_0(l+1)(l+2)}}\frac1{\hat r} \right.\nonumber \\ 
    &\times\left[\hat p_r, ...,\left[\hat p_r, e^{-\frac{\hat r}{a_0(n-2)(n-1)}} \frac1{\hat r}\left[\hat p_r, e^{-\frac{\hat r}{a_0}\frac{2n-1}{n(n-1)}}\vphantom{\frac1{\hat r}}\right.\right.\right. \nonumber\\
    &\left.\left.
    \left. \left. \times\hat r^{2n-1} \vphantom{\frac1{\hat r}}\right]\right]\cdots\right]\right]_{n-l-1}
    e^{\frac{\hat r}{na_0}}\frac1{\hat r^n}.\label{n,l}
\end{align} 
The normalization constant is found from Eq.~(\ref{normalization}), using the fact that 
\begin{equation}
    E_n-E_j = \frac{e^2}{2a_0}\left(\frac1{j^2}-\frac1{n^2}\right)=\frac{e^2(n+j)(n-j)}{2a_0n^2j^2}.
\end{equation}
This results in
\begin{align}
    &C_{n,l}  = \left(\tfrac{n\sqrt{2a_0}}{e}\right)^{n-l-1}\tfrac{(n-1)!}{l!}\sqrt{\tfrac{(n+l)!}{(2n-1)!}\tfrac{1}{(n-l-1)!}}.
\end{align}

We need to simplify the nested commutator expressions. This can again be done with differential operators, but we instead show how to do it with abstract operators. We follow the same procedure as before. We define a polynomial such that
\begin{align}
    \ket{n,l} & = C_{n,l}\tfrac{(2n-1)!l!(n-l-1)!}{(n+l)!(n-1)!}\left(-\tfrac{i\hbar}{\sqrt{2M}}\tfrac1{2\hat r}\right)^{n-l-1}\nonumber\\
    &\times L_{n-l-1}^{(2l+1)}\left(\tfrac{2\hat r}{na_0}\right)\ket{n,n-1},\label{Coulomb result}
\end{align}
and verify it is a Laguerre polynomial using induction.
Starting with $l=n-1$, we find that $L^{(2l+1)}_0\left (2\hat{r}/na_0\right )=1$. When $l=n-2$, Eq.~(\ref{n,l}) becomes
\begin{align}
    &\ket{n,n-2} = \frac{C_{n,n-2}}{\sqrt{2M}}\frac1{\hat r^{n-1}}e^{\frac{\hat r}{(n-1)a_0}}
    \left[\hat p_r, e^{-\frac{\hat r}{a_0}\frac{2n-1}{n(n-1)}}\hat r^{2n-1}\right]\nonumber\\
    &~~~\times e^{\frac{\hat r}{na_0}}\frac1{\hat r^n}\ket{n,n-1} \nonumber\\
    &~~~= -\frac{i\hbar C_{n,n-2}(2n-1)}{\sqrt{2M}(n-1)2\hat r}\left(2n-2-\frac{2\hat r}{na_0}\right)\ket{n,n-1}.
\end{align}
Using Eq.~(\ref{Coulomb result}), we find that
\begin{equation}
    L_{1}^{(2n-3)}\left(\tfrac{2\hat r}{na_0}\right)=2n-2-\frac{2\hat r}{na_0},
\end{equation}
which is also correct for this Laguerre polynomial.
To complete the proof by induction, we need to use two more recurrence relations for Laguerre polynomials,
\begin{align}
    mL_m^{(\alpha)}(x)&=(2m+\alpha-1-x)L_{m-1}^{(\alpha)}(x)\nonumber\\
    &-(m+\alpha-1)L_{m-2}^{(\alpha)}(x) \label{recurrence 2}
\end{align}
and
\begin{equation}
    L_m^{(\alpha+1)}(x)-L_{m-1}^{(\alpha+1)}(x) = L_m^{(\alpha)}(x), \label{recurrence 3}
\end{equation}
as well as the explicit formula for the Laguerre polynomials,
\begin{equation}
    L_m^{(\alpha)}(x) = \sum_{j=0}^m (-1)^j{m+\alpha\choose m-j}\frac{x^j}{j!}. \label{Laguerre coefficients}
\end{equation}

The induction step assumes that Eq.~(\ref{Coulomb result}) holds for $l$ and $l+1$, and then we need to establish that it also holds for $l-1$. Using Eq.~(\ref{n,l}), we have that
\begin{align}
    \ket{n,l-1}&=\hat{M}_{n,l-1}\ket{n,n-1} = \frac{C_{n,l-1}}{\sqrt{2M}C_{n,l}}\frac1{\hat r^l}e^{\frac{\hat r}{la_0}} \nonumber\\
    & \times \left[\hat p_r, e^{-\frac{\hat r}{la_0}}\hat r^l \hat M_{n,l} e^{-\frac{\hat r}{na_0}}\hat r^n\right] e^{\frac{\hat r}{na_0}}\frac1{\hat r^n} \ket{n,n-1}.
\end{align}
After expanding the commutator using the Leibniz product rule and simplifying, we find that this becomes
\begin{align}
    &|n,l-1\rangle=\nonumber\\
    &~~~~~~\tfrac{C_{n,l-1}}{\sqrt{2M}C_{n,l}} \left(i\hbar\hat M_{n,l}\left(\tfrac{n+l}{nla_0}-\tfrac{n+l}{\hat r}\right)+[\hat p_r,\hat M_{n,l}]\right)|n,n-1\rangle \nonumber\\
    &~~~= \tfrac{C_{n,l-1}}{\sqrt{2M}^{n-l}} \left(\tfrac{-i\hbar}{2}\right)^{n-l-1} \tfrac{(2n-1)!l!(n-l-1)!}{(n+l)!(n-1)!} \nonumber\\
    &~~~\times \left(\tfrac{i\hbar}{\hat r^{n-l-1}}L_{n-l-1}^{(2l+1)}\left(\tfrac{2\hat r}{na_0}\right)\left(\tfrac{n+l}{nla_0}-\tfrac{n+l}{\hat r}\right)\right. \nonumber\\
    &~~~+ \left.\left[\hat p_r, \tfrac1{\hat r^{n-l-1}}L_{n-l-1}^{(2l+1)}\left(\tfrac{2\hat r}{na_0}\right)\right]\right) \label{Coulomb ind}.
\end{align}
The commutator can be determined using Eq.~(\ref{Laguerre coefficients}) and evaluating the commutators term-by-term. It becomes
\begin{align}
    &\tfrac{i\hbar}{\hat r^{n-l}}\left(\tfrac{2\hat r}{na_0}L_{n-l-2}^{(2l+2)}\left(\tfrac{2\hat r}{na_0}\right) +(n-l-1)L_{n-l-1}^{(2l+1)}\left(\tfrac{2\hat r}{na_0}\right)\right)=\nonumber\\
    &~~~~~~\tfrac{i\hbar}{\hat r^{n-l}} (n+l)L_{n-l-2}^{(2l+1)}\left(\tfrac{2\hat r}{na_0}\right),
\end{align}
where the last equality used Eq.~(\ref{recurrence 1}), with $l+1/2\to 2l+1$. Substituting this result back into Eq.~(\ref{Coulomb ind}) gives
\begin{align}
    \ket{n,l-1}&= C_{n,l-1}\tfrac{(2n-1)!(l-1)!(n-l-1)!}{(n+l-1)!(n-1)!}\nonumber\\
    &\times(-1)^{n-l-1}\left(\tfrac{i\hbar}{\sqrt{2M}2\hat r}\right)^{n-l} 
    \left( 2l L_{n-l-2}^{(2l+1)}\left(\tfrac{2\hat r}{na_0}\right) \right. \nonumber\\
    & \left.+\left(\tfrac{2\hat r}{na_0}-2l\right)L_{n-l-1}^{(2l+1)}\left(\tfrac{2\hat r}{na_0}\right)\right)\ket{n,n-1}, \label{almost done}
\end{align}
after bringing in a factor of $2l/(n+l)$ from the constants out front into the parenthesis.
Using Eq.~(\ref{recurrence 2}) with $m=n-l$ and $\alpha=2l+1$ gives $(n-l)L_{n-l}^{(2l+1)}\left(2\hat r/na_0\right)=\left(2n-2\hat r/na_0\right)L_{n-l-1}^{(2l+1)}\left(2\hat r/na_0\right)-(n+l)L_{n-l-2}^{(2l+1)}\left(2\hat r/na_0\right)$. Using this fact, as well as Eq.~(\ref{recurrence 3}) multiple times, we find that
\begin{align}
    &\left( 2l L_{n-l-2}^{(2l+1)}\left(\tfrac{2\hat r}{na_0}\right) +\left(\tfrac{2\hat r}{na_0}-2l\right)L_{n-l-1}^{(2l+1)}\left(\tfrac{2\hat r}{na_0}\right)\right) \nonumber\\
    =&-(n-l)\left(L_{n-l}^{(2l+1)}\left(\tfrac{2\hat r}{na_0}\right)-2L_{n-l-1}^{(2l+1)}\left(\tfrac{2\hat r}{na_0}\right) \right.\nonumber\\
    &+\left.L_{n-l-2}^{(2l+1)}\left(\tfrac{2\hat r}{na_0}\right)\right) \nonumber\\
    =& -(n-l)\left( L_{n-l}^{(2l)}\left(\tfrac{2\hat r}{na_0}\right)-L_{n-l-1}^{(2l)}\left(\tfrac{2\hat r}{na_0}\right) \right) \nonumber\\
    = & -(n-l)L_{n-l}^{(2l-1)}\left(\tfrac{2\hat r}{na_0}\right).
\end{align}
In the first step, we replaced $(2\hat{r}/na_0)L_{n-l-1}^{(2l+1)}\left (2\hat{r}/na_0\right )$ using Eq.~(\ref{recurrence 2}) and then Eq.~(\ref{recurrence 3}) was used twice to obtain the third line and once to obtain the final result.
The final result is
\begin{align}
    \ket{n,l-1} &= C_{n,l-1}\tfrac{(2n-1)!(l-1)!(n-l)!}{(n+l-1)!(n-1)!}\nonumber\\
    &\times\left(-\tfrac{i\hbar}{\sqrt{2M}2\hat r}\right)^{n-l} L_{n-l}^{(2l-1)}\left(\tfrac{2\hat r}{na_0}\right)\ket{n,n-1},
\end{align}
which establishes the proof by induction.\par
The wavefunction of the $l=n-1$ state can then be calculated from the subsidiary condition using techniques similar to what we used before. Namely, we find that
\begin{align}
    \langle r|n,n-1\rangle &= r^{n-1} e^{-\frac{r}{na_0}}\bra{r} e^{\frac{\hat r}{na_0}}\frac1{\hat r^{n-1}}\ket{n,n-1} \nonumber\\
    &=  r^{n-1} e^{-\frac{r}{na_0}}\bra{0_r}e^{\frac{i}{\hbar}r\left (\hat{p}_r+\frac{i\hbar}{\hat{r}}\right )} e^{\frac{\hat r}{na_0}}\frac1{\hat r^{n-1}}\ket{n,n-1} \nonumber\\
    &=r^{n-1} e^{-\frac{r}{na_0}}\bra{0_r}e^{\frac{\hat r}{na_0}}\frac1{\hat r^{n-1}}\ket{n,n-1} \nonumber\\
    &= r^{n-1} e^{-\frac{r}{na_0}}\bra{0_r}\frac1{\hat r^{n-1}}\ket{n,n-1}.
\end{align}
The normalization constant is then immediately found by integration, resulting in 
\begin{equation}
    \bra{0_r}\frac1{\hat r^{n-1}}\ket{n,n-1} = \left(\frac2{na_0}\right)^{n+\frac12}\frac1{\sqrt{(2n)!}}.
\end{equation}
Combining this with the result for $C_{n,l}$ and Eq.~(\ref{Coulomb result}), gives us the final Coulomb radial wavefunction,
\begin{align}
    \psi_{n,l}(r) = \sqrt{\tfrac{(n-l-1)!}{2n(n+1)!}}\left(\tfrac2{na_0}\right)^{l+\frac32}L_{n-l-1}^{(2l+1)}\left(\tfrac{2r}{na_0}\right)r^l e^{-\frac{r}{na_0}} ,
\end{align}
which is the standard result.

\section{Approaches to include this technique in quantum instruction}

The operator Rodrigues formula approach to calculating wavefunctions has some aspects to it that are tedious, technical, and abstract. The steps leading up to the nested commutators are quite straightforward and certainly can be shown to undergraduates and graduate students alike. The determination of the special polynomial for each solution is a more complex task. Similar to how the Rodrigues polynomials are defined in terms of $n$-fold derivatives acting on the generating functions, one could just define the Rodrigues polynomials here via nested commutators and simply tell the students what the general form is without requiring them to determine the recurrence relations. This can make the approach easier to digest by students and then it has a similar level of complexity as the differential Rodrigues formula approaches where students are told the special polynomial. Graduate students, however, should be able to work with the recurrence relations, especially because the identities needed to solve all of the recurrence relations can be easily derived from the defining power series expansion for the Laguerre polynomials.

In instruction, it is probably better to spread out the content in different units, so that the material can be revisited multiple times and be absorbed more easily by the students. We would recommend covering the material for the simple harmonic oscillator in one dimension and three dimensions and for the Coulomb problem in three dimensions. The two-d examples or the Morse potential could then be assigned as exercises for the students, as could the harmonic oscillator problems in momentum space, which we did not cover in this work. They would likely need scaffolding to help students derive the recurrences if this is a goal of the homework problems.

We believe there is an elegance to these approaches, especially to how the ground-state wavefunctions are found, that students are likely to enjoy working with. We anticipate this is similar to how students prefer working with ladder operators in many different calculations for the simple harmonic oscillator instead of working with differential equations or integration. It is for this reason, that these materials are likely to be well-received by students and are worth the effort needed to work more abstractly.

\section{Summary and conclusions}

Quantum mechanics suffers from being taught primarily in the position-space representation. This is often argued to be necessary because this is the only representation where the Schr\"odinger equation is always expressed as a second-order linear differential equation. Hence, all energy eigenvalue problems are treated on the same footing. 

However, not everybody teaches quantum mechanics solely in this fashion. Most instructors will teach both the simple harmonic oscillator and the angular momentum eigenstates using an abstract, representation-independent approach. Ever since 1940, we have known how to do this for all solvable problems in quantum mechanics, but the approach has not been widely adopted. We suspect this is true primarily because using these approaches does not allow for a direct calculation of the wavefunctions in position or momentum space. This work shows how one can ameliorate such a concern and actually calculate wavefunctions without resorting to working in the position representation. Because of the importance of working with representation-independent formulations, we feel this is an important new tool that is available for instructors to use who wish to teach quantum mechanics without relying solely on the position representation.

\section{Acknowledgments}
J. R. N. was supported by the National Science Foundation under Grant No. DMR--1950502.
L.X. was supported by U.S. Department of Energy, Office of Science, Office of Advanced Scientific Computing Research (ASCR), Quantum Computing Application Teams (QCATS) program, under field work proposal number ERKJ347.
J.K.F. was supported by the National Science Foundation under Grant No. PHY-1915130 and was supported by the McDevitt bequest at Georgetown University. 
J.K.F. planned the project, including developing the steps to construct an operator-based generalization of the Rodrigues formula. J.R.N. and M.R.S. performed calculations for different Hamiltonians. L.X. served as a mentor to J.R.N.'s work. All authors contributed to writing up the work.

\section{Author Declarations}
The authors have no conflicts to disclose.

\setcounter{equation}{0}
\renewcommand{\theequation}{S\arabic{equation}}

%\section{Appendix A: Not sure we need an appendix.}
\ 
\\

%\end{document}

\newpage

\noindent\textbf{Supplemental material for Noonan, Rehman Shah, Xu, and Freericks ``Employing an operator form of the Rodrigues formula to calculate wavefunctions without differential equations.''}

\section*{Supplemental material: Summary of other problems that can be solved this way}

This strategy can be used to solve other problems as well. The approach is identical to that given in the examples we have seen already. Rather than go through these examples in full detail, we sketch how they work, providing only summary formulas. We explicitly show results for the two-dimensional isotropic oscillator and the two-dimensional Coulomb problem. The Morse potential can also be solved this way, but we do not provide any details here.

We start with the isotropic oscillator in two dimensions. We solve it in polar coordinates. The wavefunction is written in a tensor-product form $|\Psi\rangle=|\psi\rangle\otimes|m\rangle$, where $|m\rangle$ is an eigenstate of the $z$-component of angular momentum. The Hamiltonian is 
\begin{equation}
    \hat{H}=\frac{\hat{p}_x^2}{2M}+\frac{\hat{p}_y^2}{2M}+\frac{1}{2}M\omega^2(\hat{x}^2+\hat{y}^2).
\end{equation}
After separating variables and acting onto the tensor-product state, we have $\hat{H}|\Psi\rangle=\hat{H}_m|\psi\rangle\otimes|m\rangle$, with
\begin{equation}
    \hat H_m = \frac{\hat p_\rho^2}{2M} + \frac{\hbar^2(m^2-\frac14)}{2M\hat\rho^2} + \frac12M\omega^2\hat\rho^2,
\end{equation}
where $\hat{p}_\rho$ is the radial component of the momentum $\hat{p}_\rho=\tfrac{1}{\hat{\rho}}(\hat{x}\hat{p}_x+\hat{y}\hat{p}_y)-\tfrac{i\hbar}{2\hat{\rho}}$ and $\hat{\rho}^2=\hat{x}^2+\hat{y}^2$. $\hat{H}_m$ can be factorized with the lowering operator given by 
\begin{equation}
    \hat A_m = \frac1{\sqrt{2M}}\bigg(\hat p_\rho -i \bigg(\frac{-\hbar(m+\frac{1}{2})}{\hat\rho}+M\omega\hat\rho \bigg)\bigg),
\end{equation}
with energy $E_m=\hbar\omega(m+1)$. While this factorization works for all integers $m$, we focus on working with nonnegative $m$ values, because the negative values can be easily constructed from the positive ones. The auxiliary Hamiltonian for each $m$ is then $\hat A_m\hat A^\dagger_m + E_m = \hat H_{m+1}+\hbar\omega$, which is again a constant shift of the next larger constant angular momentum Hamiltonian. The intertwining relationship is $\hat H_m\hat A^\dagger_m = \hat A^\dagger_m(\hat H_{m+1}+\hbar\omega)$. The $k$th excited state of $\hat{H}_m$ satisfies $\hat H_m \hat A^\dagger_ m\hat A^\dagger_{m+1}\cdots\hat A^\dagger_{m+k-1}\ket{\phi_{m+k}} = (E_{m+k}+k\hbar\omega)\hat A^\dagger_ m\cdots\hat A^\dagger_{m+k-1}\ket{\phi_{m+k}}$, with the energy $\hbar\omega(m+2k+1)$, where $\ket{\phi_m}$ is the ground state of $\hat H_m$. The normalization constant is  $C_{m,k}=1/\sqrt{(2\hbar\omega)^k k!}$.

The similarity transformation for the raising operator is  $\hat A_m^\dagger=\frac{1}{\sqrt{2M}}e^{\frac{M\omega}{2\hbar}\hat\rho^2}\hat\rho^{-m-\frac12}\hat p_\rho \hat\rho^{m+\frac12}e^{-\frac{M\omega}{2\hbar}\hat\rho^2}$. The subsidiary condition becomes
\begin{equation}
    \hat p_\rho \hat\rho^{-m-\frac12}e^{\frac{M\omega}{2\hbar}\hat\rho^2}\ket{\phi_m}=0,
\end{equation}
which gives the state annihilated by $\hat{p}_\rho$.
The $k$th excited state of $\hat{H}_m$ is then
\begin{align}
    &\ket{\psi_{m,k}}=C_{m,k}\hat A^\dagger_ m\cdots\hat A^\dagger_{m+k-1}\ket{\phi_{m+k}} = \tfrac{1}{\sqrt{(4\hbar M\omega)^kk!}}\nonumber\\&e^{\frac{M\omega}{2\hbar}\hat\rho^2}\hat\rho^{-m-\frac12}\hat p_\rho \tfrac1{\hat\rho}\hat{p}_\rho\cdots\tfrac{1}{\hat{\rho}}\hat p_\rho\hat\rho^{m+k-1+\frac12}e^{-\frac{M\omega}{2\hbar}\hat\rho^2}\ket{\phi_{m+k}}\nonumber\\
    =&(-i)^k\tfrac{1}{\sqrt{k!}}e^{\frac{M\omega}{\hbar}\hat\rho^2}\left(\tfrac{M\omega}\hbar\hat\rho^2\right)^{-m-\frac{k}{2}}\Bigg[\tfrac{\hbar}{2M\omega\hat\rho}\tfrac{i\hat{p}_\rho}\hbar,\nonumber\\
    &\cdots,\left[\tfrac{\hbar}{2M\omega\hat\rho}\tfrac{i\hat{p}_\rho}\hbar,\left(\tfrac{M\omega}{\hbar}\hat\rho^2\right)^{m+k}e^{-\tfrac{M\omega}{\hbar}\hat\rho^2}\right]\cdots\Bigg]_k\ket{\phi_{m+k}}.
\end{align}
The Laguerre polynomial is defined as follows
\begin{align}
    &L_k ^{(m)} (\frac{M\omega}{\hbar}\hat{\rho}^2) = \frac{1}{k!}\Big(\frac{M\omega}{\hbar}\hat{\rho}^2 \Big)^{-m} e^{\tfrac{M\omega}{\hbar}\hat{\rho}^2}\nonumber\\
    &\times \Bigg[\tfrac{1}{\hat\rho}\tfrac{i\hat{p}_\rho}{2M\omega},\cdots \left[\tfrac{1}{\hat\rho}\tfrac{i\hat{p}_\rho}{2M\omega},
    \left(\tfrac{M\omega}{\hbar}\hat\rho^2\right)^{m+k}e^{-\tfrac{M\omega}{\hbar}\hat\rho^2}\right]\cdots\Bigg]_k.
    \label{eq:2dlaguerre}
\end{align}
We then need to show that $L_k ^{(m)}$ is equal to an associated Laguerre polynomial, which we will once again do by a recurrence relation. To do this, we need to show that $L_0^{m}(x)=1$ and the general recurrence relation
\begin{equation}
    kL_k ^{(m)} (x) = (m+k)L_{k-1} ^{(m)} (x) - xL_{k-1} ^{(m+1)}(x)\label{eq:recurrence4}
\end{equation}
for Laguerre polynomials. Starting with $k=0$ where the polynomial $L_0 ^{(m)}$ does not contain any commutator, Eq.~(\ref{eq:2dlaguerre}) then becomes
\begin{align}
    L_0 ^{(m)}\left (\frac{M\omega}{\hbar}\hat{\rho}^2\right ) &= \Big(\frac{M\omega}{\hbar}\hat{\rho}^2 \Big)^{-m} e^{\tfrac{M\omega}{\hbar}\hat{\rho}^2}\Big(\frac{M\omega}{\hbar}\hat{\rho}^2 \Big)^{m}e^{-\tfrac{M\omega}{\hbar}\hat\rho^2} \nonumber\\
     &= 1
\end{align}
If we evaluate the innermost commutator in Eq.~(\ref{eq:2dlaguerre}) by using Leibniz's product rule, we find that
\begin{align}
    &\left[\tfrac{1}{\hat\rho}\tfrac{i\hat{p}_\rho}{2M\omega},
    \left(\tfrac{M\omega}{\hbar}\hat\rho^2\right)^{m+k}e^{-\tfrac{M\omega}{\hbar}\hat\rho^2}\right] = \nonumber\\
    &(m+k)e^{-\tfrac{M\omega}{\hbar}\hat\rho^2} (\tfrac{M\omega}{\hbar}\hat\rho^2)^{m+k-1} - e^{-\tfrac{M\omega}{\hbar}\hat\rho^2}(\tfrac{M\omega}{\hbar}\hat\rho^2)^{m+k}
\end{align}
Thus, multiplying Eq.~(\ref{eq:2dlaguerre}) by $k$ and using it again to define the Laguerre polynomials with different indices, we find that
\begin{align}
    kL_k^{(m)}\left (\tfrac{M\omega}{\hbar}\hat{\rho}^2\right )&= (m+k)L_{k-1}^{(m)}\left (\tfrac{M\omega}{\hbar}\hat{\rho}^2\right )\nonumber\\
    &-\left (\tfrac{M\omega}{\hbar}\hat{\rho}^2\right ) L_{k-1}^{(m+1)}\left (\tfrac{M\omega}{\hbar}\hat{\rho}^2\right ),
\end{align}
which means the Laguerre polynomials defined in Eq.~(\ref{eq:2dlaguerre}) do properly satisfy the recurrence relation of the Laguerre polynomials. This then allows us to replace the iterated commutator with the Laguerre polynomial via
\begin{equation}
    \ket{\psi_{m,k}} =\sqrt{k!} \left(\tfrac{-i\sqrt{\hbar}}{\hat{\rho}\sqrt{M\omega}}\right)^k L_k^{(m)}\left(\tfrac{M\omega}\hbar\hat\rho^2\right) \ket{\phi_{m+k}}. \label{2D SHO Rodrigues}
\end{equation}
The wavefunction for the ground state can be found in the same way as we did before and is given by
\begin{equation}
    \phi_m (\rho) = \braket{\rho|\phi_m} = \rho^{m}e^{-\frac{M\omega}{2\hbar}\rho^2}\bra{0_\rho}\tfrac{1}{\hat{\rho}^{m}}\ket{\phi_m}.
\end{equation}
The term $\bra{0_\rho}\tfrac{1}{\hat{\rho}^{m}}\ket{\phi_m}$ can be found by normalizing the wavefunction
\begin{equation}
    \bra{0_\rho}\tfrac{1}{\hat{\rho}^{m}}\ket{\phi_m} = \Big(\frac{M\omega}{\hbar} \Big)^{\tfrac{m+1}{2}}\sqrt{\frac{2}{m!}}
\end{equation}
Omitting the overall phase factor $(-i)^k$, the radial wavefunction is then
\begin{align}
    \psi_{m,k}(\rho)&=\sqrt{\tfrac{k!\hbar^k}{(M\omega)^k}}\tfrac{1}{\rho}L_k^{(m)}\left(\tfrac{M\omega}\hbar\rho^2\right) \phi_{m+k}(\rho) \nonumber\\
    &=\sqrt{\tfrac{2k!}{(m+k)!}(\tfrac{M\omega}{\hbar})^{m+1}}\rho^m e^{-\frac{M\omega}{2\hbar}\rho^2}
    L_k^{(m)}\left(\tfrac{M\omega}\hbar\rho^2\right).
\end{align}
Using the principal quantum number, defined to be $n=m+2k$ and adding in the angular part of the wavefunction, we get the final wavefunction
\begin{align}
    \Psi_{n,m}(\rho,\phi) &= \tfrac{\sqrt{\big(\tfrac{n-|m|}2\big)\text{\large !}}}{\sqrt{\pi\big(\tfrac{n+|m|}2\big)\text{\large !}}}\sqrt{(\tfrac{M\omega}{\hbar})^{|m|+1}}
    L_{\frac{n-|m|}2}^{(|m|)}\left(\tfrac{M\omega}\hbar\rho^2\right)\nonumber\\
    &\times e^{-\frac{M\omega}{2\hbar}\rho^2}e^{im\phi}.
\end{align}
In this final step, we use the fact that the results for positive and negative $m$ only differ in the angular factor.\cite{spherical-translation-operator} The factor $\frac{1}{\sqrt{2\pi}}$ has been introduced to account for the normalization of $e^{im\phi}$.

The two-dimensional Coulomb problem can be solved in a similar way. The constant angular momentum Hamiltonian is
\begin{equation}
    \hat H_m = \frac{\hat p_\rho^2}{2M} + \frac{\hbar^2(m^2-\frac14)}{2M\hat\rho^2} - \frac{e^2}{\hat\rho},
\end{equation}
and the lowering operator is given by
\begin{equation}
    \hat{A}_m=\frac{1}{\sqrt{2M}}\left (\hat{p}_\rho-\frac{i\hbar}{(m+\frac{1}{2})a_0}+\frac{i\hbar(m+\frac{1}{2})}{\hat{\rho}}\right )
\end{equation}
with the energy being $E_m=-e^2/(2a_0\big(m+\frac{1}{2})^2\big)$.
As in the three-dimensional case, the $k$th auxiliary Hamiltonian for $\hat{H}_m$ is $\hat{H}_{m+k}$, and in particular, the intertwining relation is $\hat{H}_m\hat A^\dagger_{m}=\hat A^\dagger_{m}\hat{H}_{m+1}$. Thus, the eigenstates are the same form as before: $\ket{\psi_{m,k}}=C_{m,k}\hat A^\dagger_{m}\hat A^\dagger_{m+1}\cdots\hat A^\dagger_{m+k-1}\ket{\phi_{m+k}}$, with energies $E_{m+k}$, and normalization constant 
\begin{align}
    C_{m,k} &= \left(\frac{a_0\sqrt{2M}(m+k+\tfrac12)}\hbar\right)^{k} \frac{(2m+2k-1)!!}{2^k(2m-1)!!}\nonumber\\
    &\times \sqrt{\frac{(2m+k)!}{k!(2m+2k)!}}
\end{align}
The similarity transformation for the raising operator is $\hat A_m^\dagger=\frac{1}{\sqrt{2M}}e^{\frac{\hat{\rho}}{a_0(m+\frac{1}{2})}}\hat\rho^{-m-\frac12}\hat p_\rho \hat\rho^{m+\frac12}e^{-\frac{\hat{\rho}}{a_0(m+\frac{1}{2})}}$. The subsidiary condition is then
\begin{equation}
    \hat{p}_\rho\hat\rho^{-m-\frac12}e^{\frac{\hat\rho}{a_0(m+\frac12)}} \ket{\phi_m}=0.
\end{equation}
The excited states become
\begin{align}
    &\ket{\psi_{m,k}}=C_{m,k}\hat A^\dagger_ m\cdots\hat A^\dagger_{m+k-1}\ket{\phi_{m+k}} = \frac{C_{m,k}}{\sqrt{2M}^k}\times\nonumber\\
    &\left(\prod_{j=m}^{m+k-1} e^{\frac{\hat\rho}{a_0(j+\frac12)}}\hat\rho^{-j-\frac12}\hat{p}_\rho \hat\rho^{+j+\frac12}e^{-\frac{\hat\rho}{a_0(j+\frac12)}}\right)\ket{\phi_{m+k}} \nonumber\\
    &=\frac{C_{m,k}}{\sqrt{2M}^k}e^{\frac{\hat\rho}{a_0(m+\frac12)}}\hat\rho^{-m-\frac12}
    \left(\prod_{j=m}^{m+k-1} \hat{p}_\rho \frac1{\hat\rho}e^{-\frac{\hat\rho}{a_0(j+\frac12)(j+\frac32)}}\right) \nonumber\\
    &\times \hat\rho^{2m+2k+1}e^{-\frac{2\hat\rho}{a_0(m+k+\frac12)}} \hat\rho^{-m-k-\frac12}e^{\frac{\hat\rho}{a_0(m+k+\frac12)}}\ket{\phi_{m+k}} \nonumber\\
    &=\frac{C_{m,k}}{\sqrt{2M}^k}e^{\frac{\hat\rho}{a_0(m+\frac12)}}\hat\rho^{-m-\frac12} \Bigg[\hat{p}_\rho, \frac1{\hat\rho}e^{-\frac{\hat\rho}{a_0(m+\frac12)(m+\frac32)}}\Bigg[\hat{p}_\rho,\frac1{\hat\rho}\nonumber\\
    & e^{-\frac{\hat\rho}{a_0(m+\frac32)(m+\frac52)}}\cdots\Big[\hat{p}_\rho, \frac1{\hat\rho}e^{-\frac{\hat\rho}{a_0(m+k-\frac12)(m+k+\frac12)}}\hat\rho^{2m+2k+1}\nonumber\\
    & e^{-\frac{2\hat\rho}{a_0(m+k+\frac12)}}\Big]\cdots\Bigg]\Bigg]_k 
    \hat\rho^{-m-k-\frac12}e^{\frac{\hat\rho}{a_0(m+k+\frac12)}}\ket{\phi_{m+k}}. \label{eq:2DCoulombstate}
\end{align}
Now we introduce
\begin{align}
    & L_k^{(2m)}\left(\tfrac{2\hat\rho}{a_0(m+k+\frac12)}\right) = \left(-\tfrac{1}{i\hbar}\right)^{k} \tfrac{(2m+k)!}{k!(2m-1)!!(2m+2k)!!} \hat\rho^{-2m-1}\nonumber\\
    &e^{\frac{(2m+k+1)\hat\rho}{a_0(m+\frac12)(m+k+\frac12)}}\times\Bigg[\hat{p}_\rho, \frac1{\hat\rho}e^{-\frac{\hat\rho}{a_0(m+\frac12)(m+\frac32)}}\Big[\hat{p}_\rho,\frac1{\hat\rho}\nonumber\\
    &e^{-\frac{\hat\rho}{a_0(m+\frac32)(m+\frac52)}}\cdots\Big [\hat{p}_\rho, \frac1{\hat\rho}e^{-\frac{\hat\rho}{a_0(m+k-\frac12)(m+k+\frac12)}}\hat\rho^{2m+2k+1}\nonumber\\
    &e^{-\frac{2\hat\rho}{a_0(m+k+\frac12)}}\Big]\cdots\Big]\Bigg]_k .
    \label{eq:lagurre2dcolumb}
\end{align}
For $k=0$, we get $1$, and for $k=1$, we find
\begin{align}
    & L_1^{(2m)}=-\tfrac1{i\hbar}\tfrac{(2m+1)!}{(2m-1)!!(2m+2)!!}\hat\rho^{-2m-1} e^{\frac{(2m+2)\hat\rho}{a_0(m+\frac12)(m+\frac32)}}\nonumber\\
    &\left [\hat{p}_\rho, \hat\rho^{2m+2}e^{-\frac{(2m+2)\hat\rho}{a_0(m+\frac12)(m+\frac32)}}\right] = \tfrac{(2m+1)!}{(2m+2)(2m)!}\nonumber\\
    &\times \left((2m+2)-\tfrac{(2m+2)\hat\rho}{a_0(m+\frac12)(m+\frac32)}\right) = 2m+1 -\tfrac{2\hat\rho}{a_0(m+\frac32)},
\end{align}
which is the expected result for the Laguerre polynomial $L_1^{2m}\left(\frac{2\hat\rho}{a_0(m+\frac32)}\right)$. To complete the proof by induction, we need to use the recurrence relations in Eqs.~(\ref{recurrence 3}) and (\ref{eq:recurrence4}) to show that 
\begin{align}
    &L_{k+1}^{2m-2}(x)=L_{k+1}^{2m-1}(x)-L_k^{2m-1}(x)\nonumber\\
    &= L_{k+1}^{2m}(x)-L_k^{2m}(x)-(L_k^{2m}(x)-L_{k-1}^{2m}(x)) \nonumber\\
    &=L_{k+1}^{2m}(x)-2L_k^{2m}(x)+L_{k-1}^{2m}(x)\nonumber\\
    &= \tfrac1{k+1}\left((k+1+2m)L_k^{2m}(x)-xL_k^{2m+1}(x)\right)\nonumber\\
    &-2L_k^{2m}(x)+L_{k-1}^{2m}(x)\nonumber\\
    &=\tfrac1{k+1}\left((2m-1-x)L_k^{2m}(x)-(2m-1)L_{k-1}^{2m}(x)\right). 
    \label{eq:recurrence2dc}
\end{align}
Thus, we need to show that Eq.~(\ref{eq:lagurre2dcolumb}) holds for $m-1$,
\begin{align}
    & L_{k+1}^{(2m-2)}\left(\tfrac{2\hat\rho}{a_0(m+k+\frac12)}\right) = -\tfrac{(2m-1)}{i\hbar(k+1)(2m+k)}\hat\rho^{-2m+1}\nonumber\\
    &e^{\frac{(2m+k)\hat\rho}{a_0(m-\frac12)(m+k+\frac12)}}\left[\hat{p}_\rho,\hat\rho^{2m}e^{-\frac{(2m+k)\hat\rho}{a_0(m-\frac12)(m+k+\frac12)}} L_k^{(2m)} \right].
    \label{eq:2dcbeginning}
\end{align}
After expanding the commutator using the Leibniz rule and plugging the explicit formula for the Lagurre polynomials
\begin{equation}
    L_k ^{2m}(x) = \sum_{j=0}^k\frac{(-1)^j}{j!}{k+2m\choose k-j}x^j,
\end{equation}
Eq.~(\ref{eq:2dcbeginning}) becomes 
\begin{align}
    &L_{k+1}^{(2m-2)}\left(\tfrac{2\hat\rho}{a_0(m+k+\frac12)}\right) = \frac{1}{k+1}\Big(2m-1-\tfrac{2\hat\rho}{a_0(m+k+\frac12)}\Big)\nonumber\\
    &\times L_k^{(2m)}\left(\tfrac{2\hat\rho}{a_0(m+k+\frac12)}\right)- \tfrac{2m-1}{k+1}L_{k-1}^{(2m)}\Big(\tfrac{2\hat\rho}{a_0(m+k+\frac12)}\Big) \label{almost there},
\end{align}
which agrees with the recurrence relation we derived in Eq.~(\ref{eq:recurrence2dc}). Thus the Lagurre polynomial defined in Eq.~(\ref{eq:lagurre2dcolumb}) is a proper Lagurre polynomial. Then, omitting the overall phase $(-i)^k$, the wavefunction becomes
\begin{align}
    &\psi_{m,k}(\rho)=\left(\tfrac{a_0(m+k+\tfrac12)}2\right)^k \sqrt{\tfrac{(2m+2k)!k!}{(2m+k)!}} \rho^{-k}\phi_{m+k}(\rho)\nonumber\\
    &\times L_k^{2m}\left(\tfrac{2\rho}{a_0(m+k+\frac12)}\right)
\end{align}
The ground state wavefunction can be found as 
\begin{align}
    &\braket{\rho|\phi_m}= e^{-\frac{\rho}{a_0(m+\frac12)}}\rho^m \bra{0_\rho}\hat\rho^{-m}\ket{\phi_m}\nonumber\\
    &=\frac1{\sqrt{(2m+1)!}}\left(\frac2{a_0(m+\frac12)}\right)^{m+1}e^{-\frac{\rho}{a_0(m+\frac12)}}\rho^m,
\end{align}
by the same algebra as before. Hence, the radial wavefunction is
\begin{align}
    &\psi_{m,k}(\rho)= \left(\tfrac2{a_0(m+k+\frac12)}\right)^{m+1}\sqrt{\tfrac{k!}{(2m+2k+1)(2m+k)!}} \rho^m\nonumber\\
    &e^{-\frac{\rho}{a_0(m+k+\frac12)}} L_k^{2m}\left(\tfrac{2\rho}{a_0(m+k+\frac12)}\right).
\end{align}
We now define $n=|m|+k+1$, and including $\frac{e^{im\phi}}{\sqrt{2\pi}}$ for the angular wavefunction, the full wavefunction is 
\begin{align}
    &\Psi_{n,m}(\rho,\phi) = \left(\tfrac2{a_0(n-\frac12)}\right)^{|m|+1}\sqrt{\tfrac{(n-|m|-1)!}{(2n-1)(n+|m|-1)!}} \rho^{|m|}\nonumber\\
    &\times e^{-\frac{\rho}{a_0(n-\frac12)}} L_{n-|m|-1}^{|2m|}\left(\tfrac{2\rho}{a_0(n-\frac12)}\right)\frac{e^{im\phi}}{\sqrt{2\pi}}.
\end{align}


\noindent\textbf{REFERENCES}


\begin{thebibliography}{99}

\bibitem{schroedinger}
E. Schr\"odinger,  ``A Method of Determining Quantum-Mechanical Eigenvalues and Eigenfunctions," Proc. R. Irish Acad. A \textbf{46}, 9--16 (1940-41).

\color{black}
\bibitem{schroedinger2}
E. Schr\"odinger,  ``Further studies on solving eigenvalue problems by factorization,'' Proc.
R. Irish Acad. A \textbf{46}, 183--206 (1940-41).
\color{black}

\color{black}
\bibitem{schroedinger3}
E. Schr\"odinger,  ``The Factorization of the Hypergeometric Equation,''  Proc.
R. Irish Acad. A \textbf{47},  53--54, (1941-42).
\color{black}

\color{black}
\bibitem{infeld-hull}
L. Infeld,  and T. E. Hull,  ``The Factorization Method,''
Rev. Mod. Phys., \textbf{23}, 21--68, (1951).
\color{black}

\bibitem{witten}
E. Witten,  ``Dynamical Breaking of Supersymmetry,'' Nucl. Phys. B \textbf{188}, 513--554 (1981).

\color{black}
\bibitem{supplemental}
Supplemental material reference.
\color{black}

\color{black}
\bibitem{arfken}
G. B. Arfken,  H. J. Weber,  and  F. E. Harris, \textit{Mathematical methods for physicists: A comprehensive guide}, 7th ed., (Amsterdam, The Netherlands, 2013).
\color{black}

\color{black}
\bibitem{canfield}
J. Canfield,  A. Galler,  and J. K. Freericks,  ``The Laplace Method for Energy Eigenvalue Problems in Quantum Mechanics,'' Quantum Rep. \textbf{5}, 370--397 (2023).
\color{black}

\color{black}
\bibitem{ohanian}
H. C. Ohanian,  \textit{Principles of Quantum Mechanics} (Englewood Cliffs, NJ, Prentice-Hall, Inc., 1990).
\color{black}

\color{black}
\bibitem{cooper}
F. Cooper,  A. Kare,  and   U. V. Sukhatme, \textit{Supersymmetry in Quantum Mechanics} (Singapore, World Scientific, 2001). 
\color{black}

\color{black}
\bibitem{shape-invariant}
L. E. Gendenshtein,  ``Derivation of exact spectra of the Schr\"odinger equation by means of supersymmetry,''
JETP Lett., \textbf{38}, 356--359 (1983).
\color{black}

\color{black}
\bibitem{hadamard} 
E. Merzbacher,  \textit{Quantum Mechanics}, 3rd ed. (New York, John Wiley \& Sons, Inc., 1998).
\color{black}

\bibitem{spherical-translation-operator}
M. Rushka, M.  Esrick, W. N.,Mathews Jr.,  and J. K.  Freericks, ``Converting translation operators into plane polar and spherical coordinates and their use in determining quantum-mechanical wavefunctions in a representation-independent fashion,'' J. Math. Phys. \textbf{62}, 072102 (2021). 



\end{thebibliography}
\end{document}